\documentclass[aps,prb,twocolumn,superscriptaddress]{revtex4-2}
\usepackage{bm}
\usepackage{ae}
\usepackage[T1]{fontenc}
\usepackage[ansinew]{inputenc}
\usepackage{amsmath}
\usepackage{amssymb}
\usepackage{graphicx}
\usepackage{color}
\usepackage[colorlinks]{hyperref}
\usepackage{epstopdf}
\usepackage{soul}
\input{epsf}

\begin{document}

\title{Magnetoconductivity and quantum oscillations in intercalated graphite CaC$_6$ with the Fermi surface reconstructed by the uniaxial charge density wave}

\author{P. Grozi\'{c}}
\affiliation{Department of Physics, Faculty of Science, University of Zagreb, Bijeni\v{c}ka 32, Zagreb 10000, Croatia}

\author{A.M. Kadigrobov}
\affiliation{Ruhr-Universit\"{a}t Bochum, Theoretische Physik III,  Universit\"{a}tsstra\ss e 150, Bochum D-44801, Germany}

\author{Z. Rukelj}
\affiliation{Department of Physics, Faculty of Science, University of Zagreb, Bijeni\v{c}ka 32, Zagreb 10000, Croatia}

\author{I. Kup\v{c}i\'{c}}
\affiliation{Department of Physics, Faculty of Science, University of Zagreb, Bijeni\v{c}ka 32, Zagreb 10000, Croatia}


\author{D. Radi\'{c}*}
\affiliation{Department of Physics, Faculty of Science, University of Zagreb, Bijeni\v{c}ka 32, Zagreb 10000, Croatia}

\date{\today}

\begin{abstract}
We report a magnetoconductivity tensor {\boldmath $\sigma$} for the intercalated graphite CaC$_6$, in the ground state of the uniaxial charge density wave (CDW), under conditions of coherent magnetic breakdown due to strong external magnetic field $\mathbf{B}$ perpendicular to the conducting plane. The uniaxial charge density wave reconstructs initially closed Fermi surface into an open one, accompanied with formation of a pseudo-gap in the electron density of states around the Fermi energy. The magnetoconductivity tensor is calculated within the quantum density matrix and semiclassical magnetic breakdown approach focused on modification of the main, so-called "classical" contribution to magnetoconductivity by magnetic breakdown, neglecting the higher order corrections. In the presence of magnetic breakdown, in spite of open Fermi surface configuration, all classical magnetoconductivity components, the one along the CDW apex $\sigma_{xx} \sim B^{-2}$, perpendicular to the CDW apex $\sigma_{yy} \sim \mathrm{const}$, as well as the Hall conductivity $\sigma_{xy} \sim B^{-1}$, undergo strong quantum oscillations vs. inverse magnetic field. Those oscillations do not appear as a mere additive correction, but rather alter the classical result becoming an inherent part of it, turning it to essentially non-classical.
\end{abstract}

\maketitle

\section{Introduction}

The intercalated graphite compounds (GIC) \cite{Dresselhaus}, among which in this work we set our focus to CaC$_6$, have been known and studied for several decades. The intercalating atoms, mostly alkali metals from the first group, but also metals from the second and even third group, are located between graphene sheets in graphite. Beside chemical doping of $\pi$-bands of graphene sheets leading to formation of Fermi pockets, the intercalating atoms form the superlattice which introduces new periodicity on top of graphene honeycomb. Although mostly known and studied for their superconducting properties \cite{SuperconductingGIC}, the research field of certain GICs was recently widened by experimental observation of charge density waves (CDW) in CaC$_6$ \cite{Rahnejat,Shimizu}. The origin of the CDW ground state appears to be quite controversial, since the rather isotropic Fermi surface does not possess property of nesting \cite{Gruner}, fulfilling physical assumptions for paradigmatic model of the CDW instability based on it, i.e. on the Peierls instability \cite{Peierls}. In our recent paper, we proposed the model of the CDW instability in CaC$_6$ based on the topological reconstruction of the Fermi surface, from closed pockets to the open contours \cite{Anatoli-PRB2018,Petra}.\

In this paper we focus on the magnetotransport properties in such reconstructed geometry of the Fermi surface. The spacing between the open contours in the reciprocal space, due to the CDW, corresponds to the energy scale of the order of $10^2$K. Therefore, in strong magnetic field, effects of magnetic breakdown are expected to be pronounced, profoundly affecting properties of electron spectrum and transport properties. Those properties constitute the core of this paper, 
in particular the way in which magnetic breakdown modify the main, so-called "classical" contribution to magnetoconductivity. We show that, otherwise non-oscillating, classical components of magnetoconductivity tensor manifest an onset of quantum oscillations appearing due to magnetic breakdown, periodic in inverse magnetic field.\

The paper is organized in the following way: after the introduction in the first section, in the second section we present the model within which we describe the CaC$_6$ system; the third section contains calculation of the dispersion law, spectrum and wave functions under conditions of magnetic breakdown; in the fourth section we calculate the magnetoconductivity tensor and present the results; the final section contains concluding remarks and discussion.

\section{The Model}

\begin{figure*}
\centerline{\includegraphics[width=1.8\columnwidth]{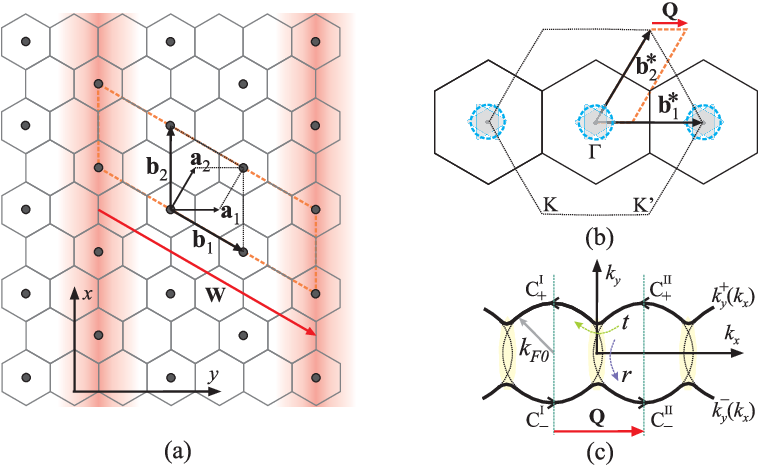}}
\caption{Schematic presentation of a 2D layer in CaC$_6$ in real and reciprocal space.  (a) In the real space, carbon atoms form a honeycomb lattice with unit vectors are $\mathbf{a}_{1,2}$ ($a \equiv \vert \mathbf{a}_{1,2} \vert \approx 2.5$\AA, the area of the cell is $A_{\mathrm{C}} \approx 5.41$\AA$^2$). Ca atoms (circles) form the hexagonal superlattice with unit vectors are $\mathbf{b}_{1,2}$ ($b \equiv \vert \mathbf{b}_{1,2} \vert = \sqrt{3}a \approx 4.32$\AA, the area of the cell is $A_{\mathrm{CaC6}} \approx 16.16$\AA$^2$). The CDW charge stripes (red-shaded along the CDW peaks) are formed along the armchair direction, creating the uniaxial periodic structure along the zig-zag direction, characterized by the vector $\mathbf{W}=3\mathbf{b}_1$ that triples the CaC$_6$ cell along $\mathbf{b}_{1}$. The new primitive cell CaC$_6\times 3$ is shown as the dashed orange rhombus. (b) In the reciprocal space, the carbon Brillouin zone (BZ) is depicted by the dashed hexagon. The Ca-superlattice, with reciprocal unit vectors $\mathbf{b}_{1,2}^*$ ($b^* \equiv \vert \mathbf{b}_{1,2}^* \vert \approx 1.68$\AA$^{-1}$), folds the carbon BZ to a three times smaller CaC$_6$ BZ (solid hexagon). All 6 Fermi pockets, from carbon K and K' points, fall into the $\Gamma$ point (shaded), approximated by a circle of the same area $S_{F0}$ depicted by the dashed blue circle. The chemical doping of $\xi \approx 0.2$ electrons per carbon atom \cite{Rahnejat} is related to the area of the Fermi pocket $S_{0}=2\pi^2\xi/A_{\mathrm{CaC6}}\approx 0.244$\AA$^{-2}$, which gives an average Fermi wave number $k_{F0}\approx 0.28$\AA$^{-1}$. The CDW potential, with the wave vector $\mathbf{Q} \parallel \mathbf{b}_{1}^*$ of periodicity $Q=b^*/3 \approx 0.56$\AA$^{-1}$, folds the CaC$_6$ BZ, bringing the FSs into touch (or slight overlap). The corresponding unit cell in reciprocal space is marked by dashed orange rhombus. (c) The Fermi surface reconstructed by the CDW potential, forming the open sheets in $k_x$-direction. Arrows show the direction of semiclassical motion of electrons in external magnetic field $\mathbf{B}$ perpendicular to the sample. Magnetic breakdown (MB) affects the semiclassical motion causing electrons to pass through the MB-junction (shaded) with probability amplitude $t(B)$, or get reflected from it with probability amplitude $r(B)$, $C_\pm^{\mathrm{I,II}}$ are coefficients denoting the branches of semiclassical wave functions corresponding to trajectories $k_y^{\pm}(k_x; \varepsilon)$.} 
\label{Fig_CaC6structure}
\end{figure*}

We model the CaC$_6$ system as a 2D graphene sheet, chemically doped by electrons from intercalating atoms to provide a finite electron pocket at the Fermi surface. The underlying Ca-lattice is of hexagonal symmetry, comprising three carbon primitive cells into the new CaC$_6$ supercell. This periodic potential folds the original carbon Brillouin zone (BZ) to the new one, three times smaller, with Fermi pockets, originally located at 6 graphene K and K' points, falling to the center of the new zone ($\Gamma$ point). The Fermi surface, for the matter of presentation and simplicity, can be approximated with the 6-fold degenerate circle, while the details of the shape of the Fermi pockets can be addressed in the conductivity calculations as parameters appearing as effective carrier concentrations. The uniaxial CDW is formed with peaks along the graphene armchair direction, with periodicity that triples the CaC$_6$ cell. This further, uniaxial reduction of the BZ brings the Fermi pockets to touching or slight overlap, leading in turn to the reconstruction of the Fermi surface due to finite CDW order parameter acting as the gap parameter in electron spectrum. The Fermi surface is topologically reconstructed: from the closed pockets, it is turned into set of open sheets. To study the magnetoconductivity, the system is put into an external homogeneous magnetic field $\mathbf{B}$, perpendicular to the sample plane. The configuration of the real and reciprocal space is schematically shown in Fig. \ref{Fig_CaC6structure}.\\    

%
%

The zero-field electron spectrum in the CDW groundstate attains the well known form (see Ref. \cite{Petra} for details)
%
\begin{eqnarray}
E_{\pm}(\mathbf{k}) &=& \frac{1}{2} \left[ \varepsilon (\mathbf{k}-\tfrac{\mathbf{Q}}{2})+\varepsilon ( \mathbf{k}+\tfrac{\mathbf{Q}}{2} ) \, \pm \right. \nonumber\\
&& \left. \sqrt{\left( \varepsilon ( \mathbf{k}-\tfrac{\mathbf{Q}}{2} ) - \varepsilon ( \mathbf{k}+\tfrac{\mathbf{Q}}{2} ) \right)^2+4\Delta^2} \right],
\label{BandsB0}
\end{eqnarray}
%
where $\varepsilon(\mathbf{k})=\hbar v_F\vert\mathbf{k}\vert$ is the initial electron spectrum - the Dirac-like electron dispersion with the Fermi velocity $v_F$ and electron wave vector $\mathbf{k}=(k_x,k_y)$, while $\mathbf{Q}=(Q,0)$ is the CDW wave vector.
Here, the origin of the reciprocal space is conveniently chosen at the crossing point of the initial electron bands (the edge of the reconstructed BZ). Due to finite CDW order parameter, $\Delta$, the degeneracy in the band crossing region is lifted, leading to the reconstruction of the FS, as shown schematically in Fig. \ref{Fig_CaC6structure}c.

\section{Electron spectrum and wave functions in finite magnetic field}

To obtain the electron spectrum in external magnetic field $\mathbf{B}$ perpendicular to the sample, under conditions of magnetic breakdown (MB), we utilize a semiclassical technique based on the Lifshitz-Onsager Hamiltonian \cite{Onsager,Lifshitz} which describes semiclassical motion of electrons between the MB regions. The necessary assumption, required to formulate magnetic breakdown problem beyond the mere perturbative contribution of magnetic field, is that the field is strong enough to provide the Larmor radius of electron motion much smaller that the mean free path of scattering on impurities. The further assumption is the absence of dislocation fields, required to provide conditions for so-called {\it coherent magnetic breakdown} \cite{Kaganov-Slutskin} which is in the focus of this paper.
The limit of so-called {\it stohastic magnetic breakdown} \cite{Falicov-Stachowiak} is not a subject of this work.\\

Choosing the Landau gauge of the vector potential $\mathbf{A}=(0,Bx,0)$, the Lifshitz-Onsager Hamiltonian leads to the Schr\"{o}dinger equation in the reciprocal space
%
\begin{eqnarray}
\varepsilon_{\nu}\left( k_x, K_y - i \frac{b^2_B}{\hbar^2} \frac{\mathrm{d} }{\mathrm{d} k_x} \right) G_{\nu}(k_x,K_y)=\varepsilon \,  G_{\nu}(k_x,K_y),\nonumber\\
\label{LO_Schrod}
\end{eqnarray}
%
where $\varepsilon_\nu(k_x,k_y)$ is the initial electron dispersion shifted in the reciprocal space to the position corresponding to trajectories I, II with branches $\pm$ (see Fig. \ref{Fig_CaC6structure}c), $b_B=\sqrt{e\hbar B}$ is the "magnetic length" (in momentum space) for electron with charge $-e$, $\hbar K_y$ is the conserved generalized momentum of the semiclassical motion of electron in the used gauge, $\varepsilon$ is the eigenvalue of energy. The semiclassical eigenfunctions are    
%
\begin{eqnarray}
G_{\pm} (k_x,K_y) = \frac{C_{\pm}}{\sqrt{|v_y^{\pm}|}} \exp \left[i\frac{\hbar^2}{b_B^2} \int^{k_x} \left(k_y^{\pm}( k_x^{\prime};\varepsilon_F) \right. \right. \nonumber\\
\left. \left. - K_y \right) \mathrm{d} k_x^{\prime} \right]
\label{WF}
\end{eqnarray}
%
analogous for both regions I, II (we omit writing these indices for simplicity here), where $C_{\pm}$ are the corresponding coefficients, $v_y^{\pm} \equiv v_y(k_x;k_y^\pm(k_x,\varepsilon))$ are the group velocity components of $\mathbf{v}=\tfrac{1}{\hbar}\nabla_{\mathbf{k}} \varepsilon(\mathbf{k})$ along electron semiclassical trajectories at energy $\varepsilon$. Coefficients $C_\pm$ are found by matching the wave functions (\ref{WF}) at the MB points. The integral in the exponent is the semiclassical phase (area enclosed by the trajectory in the reciprocal space, i.e. the semiclassical action) with lower limit determined by the starting point of the trajectory along the $k_y^{\pm}( k_x;\varepsilon_F)$ at the Fermi energy $\varepsilon_F$ in each region I, II.
Note that these trajectories in the presented procedure are found from the equation  $\varepsilon_\pm(k_x,k_y) = \varepsilon_F$, i.e. from the initial electron dispersion with gap parameter $\Delta$ neglected (dotted trajectories in Fig. \ref{Fig_CaC6structure}c). Therefore the solutions are valid far from the MB-regions. In the considered case this dependence is simply $k_y^{\pm}( k_x;\varepsilon)=\pm \sqrt{(\varepsilon/\hbar v_F)^2 - k_x^2}$.\\

\begin{figure}
\centerline{\includegraphics[width=\columnwidth]{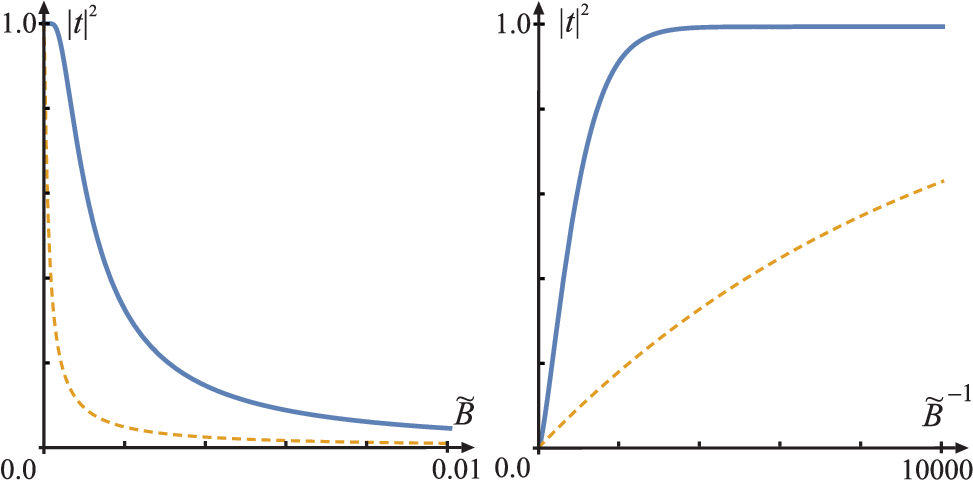}}
\caption{The MB parameter $|t(B)|^2$ according to Eq. (\ref{MBprobab-t}), where B is scaled in terms of magnetic energy, i.e. $\widetilde{B} \equiv \hbar\omega_c/\varepsilon_F$. Cyclotron frequency is taken at the Fermi energy, while the gap parameter is set to the typical order of magnitude of $\Delta/\varepsilon_F=0.01$. Dotted curve is the Blount's result \cite{Blount} plotted for comparison. Left panel shows dependence on $\widetilde{B}$, the right panel on $\widetilde{B}^{-1}$.}
\label{Fig_t2}
\end{figure}

The semiclassical solutions $G_\pm^{\mathrm{I,II}}$ in regions I and II (see Fig. \ref{Fig_CaC6structure}c), characterized by coefficients $C_{\pm}^{\mathrm{I,II}}$, are connected in the "MB-junction" by the MB-scattering matrix that relates pairs of incoming and outgoing electron waves 
%
\begin{eqnarray}
\begin{pmatrix}
C_-^\mathrm{I}\\
C_+^\mathrm{II}
\end{pmatrix}=
\mathrm{e}^{i\theta}
\begin{pmatrix}
t & r\\
-r^{\ast} & t^{\ast}
\end{pmatrix}
\begin{pmatrix}
C_-^\mathrm{II}\\
C_+^\mathrm{I}
\end{pmatrix}.
\label{MB_matrix}
\end{eqnarray}
%
Here $t(B)$ and $r(B)$, fulfilling the unitarity condition $|t|^2+|r|^2=1$, are the complex probability amplitudes for electron to pass through the MB-region and to get reflected on it, respectively, while $\theta$ is the phase determined by the problem-specific boundary conditions. It has been shown \cite{EPJB,PhysicaB,Voroncov,Fortin} that, in the configuration originating form the very slight overlap of semiclassical trajectories, the probability of passing through the MB-region is
%
\begin{eqnarray}
|t(B)|^2\approx 1- \exp{\left[-\frac{\Delta^2}{\hbar \omega_c\varepsilon_F} \sqrt[3]{\frac{\varepsilon_F}{\hbar \omega_c}} \, \right]},
\label{MBprobab-t}
\end{eqnarray}
%
where $\omega_c \equiv eB/m^*$ is the cyclotron frequency for electron with effective cyclotron mass $m^*$ ($\hbar \omega_c$ is then "magnetic energy").
Here, the latter physical quantities are just introduced as terms while their specific forms and energy dependence will be elaborated later, in the next section. In our consideration the limit $|t|=1$ accounts for the total transparency of the "MB junction" with zero reflection, i.e. the absence of (or negligible) magnetic breakdown. Regime $|t(B)|<1$ accounts for finite magnetic breakdown, the activation of over-gap tunneling assisted by magnetic field. It is worth mentioning that, despite its name, there is no typical breakdown with some finite threshold field, but rather exponential activation of the tunneling at any finite field.
One immediately notices that the exponent in Eq. (\ref{MBprobab-t}) has an additional large factor, i.e. the third root of ratio of the Fermi energy and magnetic energy, compared with the standard Blount's result \cite{Blount} obtained for arbitrary large overlap of trajectories. It is result of the peculiar band topology in the reconstruction region. 
Dependence of $|t(B)|^2$ is shown in Fig. \ref{Fig_t2}.
%
%

Periodic boundary conditions, imposed upon the semiclassical solutions by the CDW, i.e. $G_\pm(k_x,K_y)=G_\pm(k_x+Q,K_y)$, yield two additional relations between four coefficients $C_\pm^{\mathrm{I,II}}$.
They constitute, together with (\ref{MB_matrix}), a homogeneous system of two algebraic equations for two unknowns $C_+^{\mathrm{I}}$ and $C_-^{\mathrm{I}}$, i.e.
%
\begin{eqnarray}
\begin{pmatrix}
C_-^\mathrm{I}\\
C_+^\mathrm{I}
\end{pmatrix}=
\mathrm{e}^{i\theta}
\begin{pmatrix}
t & r\\
-r^{\ast} & t^{\ast}
\end{pmatrix}
\begin{pmatrix}
C_-^\mathrm{I} \exp{\left[-i\frac{\hbar^2}{b_B^2} \left( S_- + QK_y \right) \right]}\\
C_+^\mathrm{I} \exp{\left[-i\frac{\hbar^2}{b_B^2} \left( S_+ - QK_y \right) \right]}
\end{pmatrix}.\nonumber\\
\label{MB_matrix_periodicity}
\end{eqnarray}
%
Here, $S_+(\varepsilon)=\int_0^Q k_y^+(k_x;\varepsilon)dk_x$, $S_-(\varepsilon)= \int_Q^0 k_y^-(k_x;\varepsilon)dk_x$ are the semiclassical actions along the corresponding electron trajectories.
The determinant of that system, taken at arbitrary energy $\epsilon$, reads
%
\begin{eqnarray}
D(\varepsilon, K_y)=\cos \left( \frac{ \hbar^2 S_0(\varepsilon)}{2 b_B^2} + \theta \right) -|t|\cos\left( \frac{ \hbar^2 Q K_y}{b_B^2} +\mu \right),\nonumber\\
\label{D}
\end{eqnarray}
%
where $S_0(\varepsilon)=S_+(\varepsilon)+S_-(\varepsilon)$ is area of reciprocal space enclosed by electron trajectory (dotted circle in Fig. \ref{Fig_CaC6structure}c), assuming mirror symmetry of $k_y^+(k_x)$ and $k_y^-(k_x)$ along the $k_x$-axis, i.e. $S_+ = S_-$.
Phase $\mu$, appearing from $t=|t|\exp(i\mu)$, is, along with the phase $\theta$, determined by the boundary conditions of the problem specific for the particular MB configuration depending on magnetic field. They are obtained by matching the semiclassical solution to the asymptotic form of exact quantum-mechanical solution within the MB region \cite{EPJB,PhysicaB}.
Although in electron spectrum and some related quantities these phases indeed play a role, in the problem of magnetoconductivity that we consider they appear to be irrelevant. We will keep them in this section for the sake of "bookkeeping", but in the calculation of magnetoconductivity they will be omitted since they are integrated out anyway in expansions of periodic functions. Generally speaking, in problems involving magnetic breakdown determining electron spectrum can be very challenging, if possible at all. In some cases, spectrum can have very complicated structure, for example possessing fractal properties. In that respect to deal with number of quantities depending on electron spectrum, such as magnetoconductivity that we will explore in the next section, methods to utilize determinant $D(\varepsilon, K_y)$ Eq. (\ref{D}) instead of electron spectrum were developed (see for example Ref. \cite{Kaganov-Slutskin}). Determinant $D$ is function of electron energy $\varepsilon$ and its conserved momentum $K_y$ yielding the corresponding partial derivatives
%
\begin{eqnarray}
\frac{\partial D}{\partial K_y} \Big|_{\varepsilon} = |t|\frac{\hbar^2Q}{b_B^2} \sin \left( \frac{\hbar^2QK_y}{b_B^2} +\mu \right), \nonumber\\
\frac{\partial D}{\partial \varepsilon} \Big|_{K_y} = -\frac{\pi \varepsilon}{v_F^2 b_B^2} \sin \left( \frac{\pi \varepsilon^2}{2v_F^2 b_B^2} +\theta \right),
\label{D_partial_derivatives}
\end{eqnarray}
%
which will be used in the section that follows.

Electron spectrum $\varepsilon_n(K_y)$ is determined from the dispersion equation 
%
\begin{eqnarray}
D(\varepsilon, K_y)=0.
\label{D=0}
\end{eqnarray}
%

For example, in large enough field to produce a very strong magnetic breakdown $|t(B)| \rightarrow 0$, $|r(B)| \rightarrow 1$,  the dominant electron motion is along the closed orbits (dotted circles in Fig. \ref{Fig_CaC6structure}c) due the maximized over-gap tunneling between open trajectories $k_y^+(k_x)$ and $k_y^-(k_x)$. Orbital effect of magnetic field is then a mere Landau quantization of closed orbits. Dispersion law reduces to $\cos(\hbar^2 S_0(\varepsilon)/2b_B^2+\theta)=0$ which, assuming the initial spectrum $\varepsilon=\hbar v_F |\mathbf{k}|$ and $S_0=\pi |\mathbf{k}|^2$, yields the Landau-quantized spectrum in magnetic field $\varepsilon_n = \pm  v_F b_B \sqrt{2(n+1/2-\theta/\pi)}$, $n=0,1,2,...$. In contrast to the monolayer graphene, where the nontrivial geometric (Berry) phase $\phi_B = \pi$ appears leading to spectrum with Landau level $\varepsilon_{n=0}$ at zero energy, in graphite the geometric phase is trivial \cite{Mikitik}.
The $|t|=0$ case in graphite yields $\theta=0$ \cite{EPJB,PhysicaB}, finally resulting in spectrum $\varepsilon_n= \pm v_F \sqrt{2e\hbar B (n+1/2)}$, which we adopt in our consideration although, for the sake of modelling, we use the two-dimensional formalism.

\begin{figure}
\centerline{\includegraphics[width=\columnwidth]{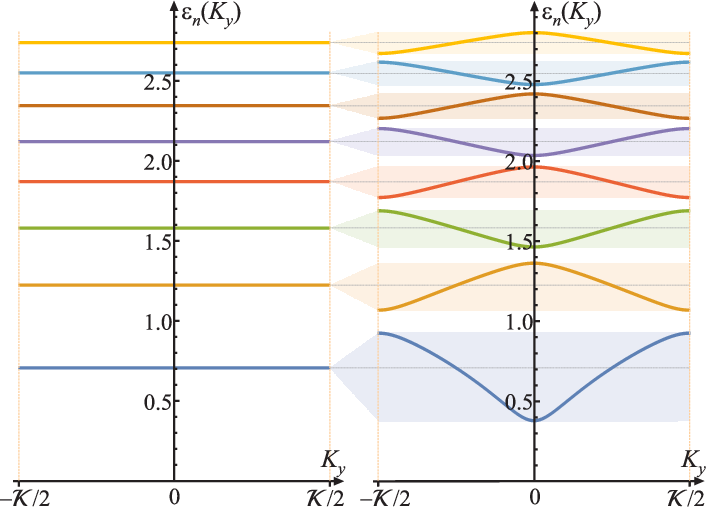}}
\caption{Spectrum (\ref{Spectrum}) in the nondispersive $|t| \rightarrow 0$ limit, essentially consisting of Landau levels (left panel), and for $|t|=0.9$ where magnetic bands are formed due to magnetic breakdown (right panel). Energy $\varepsilon_n(K_y)$, $n=0,1,...,7$ is scaled to $v_F\sqrt{2e\hbar B}$ and displayed along the "magnetic zone" of width $\mathcal{K}= 2\pi b_B^2/\hbar^2 Q = 2\pi eB/\hbar Q$. For the matter of presentation we set phases $\theta=\mu=0$.}
\label{Fig_Spectrum}
\end{figure}

The spectrum for arbitrary $|t(B)|$ can be obtained in the closed form for the considered case, reading
%
\begin{eqnarray}
\hspace{-3mm} \varepsilon_n(K_y) &=& \pm v_F \sqrt{2e\hbar B} \left[ n+\frac{1}{2}\left( 1-(-1)^n \right) -\frac{\theta}{\pi} \right. \nonumber\\
 &+& \left. \frac{(-1)^n}{\pi} \arccos{\left( |t| \cos{\left( \frac{\hbar^2 Q K_y}{b_B^2} + \mu \right)} \right)} \right]^\frac{1}{2}
\label{Spectrum}
\end{eqnarray}
%
for $n=0,1,2,...$, shown in Fig. \ref{Fig_Spectrum}.

%
%

Besides the dispersion law and spectrum, the system (\ref{MB_matrix_periodicity}) also determines relation between coefficients, i.e.
%
\begin{eqnarray}
C_- &=& \frac{|r|\exp{\left[i(\phi -\kappa_y + \theta)\right]}}{1-|t|\exp{\left[i(\phi +\kappa_y + \theta)\right]}} C_+,
\label{CoefficientsRelation}
\end{eqnarray}
%
where $\kappa_y \equiv \hbar^2 Q K_y / b_B^2$ and $\phi \equiv S_+ = S_-$.

The complete semiclassical wave function is
%
\begin{eqnarray}
G_{\eta}(k_x) &=& \frac{C_+}{\sqrt{|v_y^+|}} \exp{\left[i\frac{\hbar^2}{b_B^2}\int_0^{k_x} \left(k_y^+(k_x') - K_y \right)\mathrm{d}k_x' \right]} \nonumber\\
&+& \frac{C_-}{\sqrt{|v_y^-|}} \exp{\left[i\frac{\hbar^2}{b_B^2}\int_Q^{k_x} \left(k_y^-(k_x') - K_y \right)\mathrm{d}k_x' \right]}. \nonumber\\
\label{Semiclassical_WF}
\end{eqnarray}
%
Constants $C_+$ and $C_-$ are also related by the normalization condition of the wave function $\langle G_\eta(k_x)\mid G_\eta(k_x) \rangle =1$, where $\eta=\lbrace K_y,n \rbrace$ is set of all good quantum numbers. After neglecting the fast-oscillating cross-terms, sum of their absolute squares reduces to
%
\begin{eqnarray}
|C_+|^2+|C_-|^2 = \left( \frac{L_x}{2\pi} \int_0^Q{\frac{\mathrm{d}k_x}{|v_y(k_x)|} } \right)^{-1},
\label{NormalizationSum_I}
\end{eqnarray}
%
where $L_x$ is the length of the sample in $x$-direction. Here we used the fact that due to symmetry of the $\pm$ trajectories, the velocity components are equal by the absolute value, i.e. $|v_y| \equiv |v_y^+| = |v_y^-|$. For further procedures, it turns convenient to normalize the wave function to period of the cyclotron motion around the semiclassical orbit. Assuming this motion to be governed by the Lorentz force, i.e. $\hbar \mathrm{d}\mathbf{k}/\mathrm{d}t = -e \mathbf{v} \times \mathbf{B}$, and from there substituting $\mathrm{d}k_x = -\tfrac{1}{\hbar}eBv_y \mathrm{d}t$ into Eq. (\ref{NormalizationSum_I}), we obtain 
%
\begin{eqnarray}
|C_+|^2+|C_-|^2 = \frac{2\pi\hbar}{eBL_x} \frac{2}{T(\varepsilon)}.
\label{NormalizationSum_II}
\end{eqnarray}
%
Here $T(\varepsilon)$ is period of electron motion around circular semiclassical trajectory at energy $\varepsilon$, i.e. $T/2$ corresponds to the integral over $k_x$ from 0 to $Q$ in Eq. (\ref{NormalizationSum_I}). $T$ is related to the cyclotron frequency $\omega_c$ in the standard way, $T=2\pi/\omega_c$. $|C_+|^2$ and $|C_-|^2$ are determined by the system of equations (\ref{CoefficientsRelation},\ref{NormalizationSum_II}).

\section{Magnetoconductivity}

The magnetoconductivity tensor of a 2D system is obtained using the quantum density matrix formalism up to the linear correction to the equilibrium conditions (see Ref. \cite{Barbara} for the derivation details), with the general form of it reading
%
\begin{eqnarray}
\sigma_{\alpha \beta} = -\frac{2\zeta e^2}{L_x L_y} \sum_{\eta,\eta'}{\frac{\langle \eta \vert \hat v_\alpha \vert \eta' \rangle \langle \eta' \vert \hat v_\beta \vert \eta \rangle}{\tfrac{i}{\hbar}\left( \varepsilon_{\eta'} - \varepsilon_{\eta} \right)+\tfrac{1}{\tau_0}} \, \frac{\mathrm{d}f(\varepsilon)}{\mathrm{d}\varepsilon}\Big|_{\varepsilon=\varepsilon_\eta}}, \nonumber\\
\label{Sigma_general}
\end{eqnarray}
%
where $\alpha,\beta \in \lbrace x,y \rbrace$ account for directions along the real space (see Fig. \ref{Fig_CaC6structure}a) containing a sample of the size $L_x$ by $L_y$, factor 2 accounts for the spin degeneracy and $\zeta=6$ for the CaC$_6$ degeneracy (2 graphene valleys and tripling of the unit cell). Function $f(\varepsilon)$ 
%
%
is the Fermi distribution function at temperature $T$. We denote an operator by "hat" upon it, while $\eta$ denotes a complete set of all quantum numbers, in our problem $\lbrace n,K_y \rbrace$. In this expression, $\tau_0$ is the relaxation time due to electron scattering on impurities. We emphasize again that results to be presented are derived in the limit in which the impurity scattering rate is much smaller comparing to the cyclotron frequency, i.e. $\omega_c \gg \tau_0^{-1}$. The next important physical scale, within the regime of coherent magnetic breakdown, is the width of magnetic bands $W(B)$ compared to the broadening of level due to scattering on impurities. $W(B)$ depends on magnetic field, essentially being controlled by the tunneling probability amplitude $|t(B)|$ (\ref{MBprobab-t}) through Eq. (\ref{Spectrum}). In the limit $W(B) \ll \hbar \tau_0^{-1}$ the structure of magnetic bands and accompanying interference effects disappear and physics reduces to the merely Landau level physics. On the contrary, in the opposite limit  $W(B) \gg \hbar \tau_0^{-1}$, we expect the full-scale MB effects to be pronounced, these being the goal of this paper.

The $2 \times 2$ magnetoconductivity is anisotropic. It contains: (1) $\sigma_{xx}$ component along the CDW peaks in real space and perpendicular to the open electron trajectories in the reciprocal space; (2) $\sigma_{yy}$ component along the CDW periodicity direction in real space and along the open electron trajectories in the reciprocal space; (3) $\sigma_{xy}=-\sigma_{yx}$ are the Hall conductivity components.

\subsection{Diagonal magnetoconductivity $\sigma_{xx}$}

The diagonal magnetoconductivity along $x$-direction has to be calculated directly from Eq. (\ref{Sigma_general}) by evaluating the matrix element $\langle \eta |\hat v_x| \eta' \rangle$ due to vanishing semiclassical group velocity along that direction at the apex of the corresponding trajectory (see Fig. \ref{Fig_CaC6structure}c, $k_y$-direction). Using the above-mentioned expression for the Lorentz force, $\hbar \dot k_y =ev_x B$, and equation of motion for the momentum operator $\dot{\hat{k}}_y = \tfrac{i}{\hbar} \left[ \hat{\mathcal{H}},\hat{k}_y \right]$, we obtain $\hat v_x = \tfrac{i}{eB} \left[ \hat{\mathcal{H}},\hat{k}_y \right]$, where $\hat{\mathcal{H}}$ is Hamiltonian of the system with eigenvectors $| \eta \rangle$ and corresponding eigenvalues $\varepsilon_\eta$ \cite{Kaganov-Slutskin,Kaganov-Peschansky}. The sought for matrix element can be directly evaluated, i.e. $\langle \eta |\hat v_x| \eta' \rangle = \tfrac{i}{eB}(\varepsilon_\eta-\varepsilon_\eta')\langle \eta |\hat k_y| \eta' \rangle$. Inserting this expression in Eq. (\ref{Sigma_general}) the $\eta = \eta'$ contributions in double summation vanish. For $\eta \ne \eta'$ we expand fraction $\left( \tfrac{i}{\hbar} (\varepsilon_{\eta'} - \varepsilon_{\eta}) + \tau_0^{-1} \right)^{-1}$ under assumption $| \varepsilon_\eta-\varepsilon_{\eta'} | \gg \hbar \tau_0^{-1}$ up to the second term in Taylor series. After performing one summation over the complete set $|\eta'\rangle$ and using $\langle \eta |\hat k_y| \eta \rangle = 0$ for a symmetric trajectory, Eq. (\ref{Sigma_general}) reduces to
%
\begin{eqnarray}
\sigma_{xx} = -\frac{2\zeta e^2 \hbar^4}{L_x L_y \tau_0 b_B^4} \sum_{\eta}{\langle \eta \vert \hat{k}_y^2 \vert \eta \rangle \, \frac{\mathrm{d}f(\varepsilon)}{\mathrm{d}\varepsilon}\Big|_{\varepsilon_\eta}}.
\label{Sigma_xx1}
\end{eqnarray}
%
To evaluate the matrix element $\langle \eta \vert \hat{k}_y^2 \vert \eta \rangle$, we use semiclassical wave functions (\ref{Semiclassical_WF}), i.e. $|\eta\rangle = G_{n,K_y}(k_x)$, yielding
%
\begin{eqnarray}
\langle \eta \vert \hat{k}_y^2 \vert \eta \rangle &=& \frac{L_x}{2\pi}\int_0^Q \mathrm{d}k_x\frac{\left( k_y^+(k_x,\varepsilon_n(K_y)) \right)^2}{|v_y(k_x,\varepsilon_n(K_y))|} \nonumber\\
&\times & \left( \vert C_+(K_y,\varepsilon_n(K_y)) \vert^2 + \vert C_-(K_y,\varepsilon_n(K_y)) \vert^2 \right). \nonumber\\ 
\label{ky2_MatrixElement}
\end{eqnarray}
%
Fraction in Eq. (\ref{ky2_MatrixElement}) can be further simplified using relations $\varepsilon(k_x,k_y)=\hbar v_F \sqrt{k_x^2+k_y^2}$ and $\hbar v_y = \partial \varepsilon / \partial k_y$, i.e. $k_y^2/|v_y| = m^*(\varepsilon) k_y /\hbar$, where $m^*(\varepsilon)$ is effective cyclotron mass introduced in the previous section (although not everywhere written explicitly for the sake of convenience, $m^*(\varepsilon)$ and $\omega_c(\varepsilon)$ are functions of energy $\varepsilon$ and are treated as such in further calculations). It is a well-known physical quantity, i.e. $m^*(\varepsilon) \equiv \tfrac{\hbar^2}{2\pi}\mathrm{d}S(\varepsilon)/\mathrm{d}\varepsilon$, where $S(\varepsilon)$ is the area enclosed by electron trajectory in the reciprocal space at energy $\varepsilon$. In the case of graphene and graphite it is $m^*(\varepsilon) = \varepsilon /v_F^2$ for low enough energy to preserve linear dispersion. 
Inserting Eq. (\ref{ky2_MatrixElement}) into Eq. (\ref{Sigma_xx1}) and changing the variable $\varepsilon_n \rightarrow \varepsilon$ by inserting the integral over delta function $\int \mathrm{d}\varepsilon \, \delta(\varepsilon - \varepsilon_n)$ we obtain
%
\begin{eqnarray}
\sigma_{xx} &=& -\frac{2\zeta e^2 \hbar^4}{L_x L_y \tau_0 b_B^4} \int \mathrm{d}\varepsilon  \, \left( \frac{m^*(\varepsilon)}{\hbar} \frac{\mathrm{d}f(\varepsilon)}{\mathrm{d}\varepsilon} \right. \nonumber\\
&\times & \frac{L_x}{2\pi}\int_0^Q \mathrm{d}k_x k_y^+(k_x,\varepsilon) \nonumber\\
&\times & \frac{L_y}{2\pi} \int_0^{K_y^{m}} \mathrm{d}K_y \left\lbrace \, \left( \vert C_+(K_y,\varepsilon) \vert^2 + \vert C_-(K_y,\varepsilon) \vert^2 \right) \right.  \nonumber\\
&\times & \left. \sum_n \delta(\varepsilon - \varepsilon_n(K_y)) \left. \right\rbrace \right) .
\label{Sigma_xx2}
\end{eqnarray}
%

The integral over $k_x$ (the second row of Eq. (\ref{Sigma_xx2})) is approximately evaluated to the half-size of electron pocket at energy $\varepsilon$, i.e. $\int_0^Q k_y^+(k_x,\varepsilon)\mathrm{d}k_x \approx S_0(\varepsilon)/2$. At energy equal to Fermi, $\varepsilon=\varepsilon_F$, it gives the size of the Fermi surface which determines the number of carriers per spin projection.
The integral over $K_y$ (the third and fourth row of Eq. (\ref{Sigma_xx2})) is calculated taking into account the normalization condition (\ref{NormalizationSum_II}), which evaluates to $|C_+|^2+|C_-|^2=2\hbar/(L_x m^*(\varepsilon))$, and utilizing the well-known decomposition of delta function over zero-points of its argument, $\delta(g(x))= \sum_l \delta(x-x_l)/|g'(x_l)|$, yielding the identity
%
\begin{eqnarray}
\sum_n{\delta(\varepsilon-\varepsilon_n(K_y))}=\Big| \frac{\partial D}{\partial \varepsilon} \Big| \delta(D(\varepsilon,K_y)). 
\label{identity-delta}
\end{eqnarray}
%
Partial derivative is known from Eq. (\ref{D_partial_derivatives}), while delta function is expanded in Fourier series in the following way
%
\begin{eqnarray}
\delta(D(\varepsilon,K_y))=\sum_l{A_l(\varepsilon) \exp{\left(il \frac{\hbar^2QK_y}{b_B^2} \right)}}. 
\label{delta-expansion}
\end{eqnarray}
%
Contribution from $l=0$, i.e. $A_0(\varepsilon)$, is what we call the \textit{main contribution} or often the "classical result", while $l \ne 0$ contributions represent the fast-oscillating \textit{corrections} to it. In our consideration, we are interested in effects of magnetic breakdown to the main contribution and eventual modification of otherwise non-oscillating classical result. Therefore, in expression (\ref{delta-expansion}) we keep only the $l=0$ contribution, i.e.
%
\begin{eqnarray}
A_0(\varepsilon) &=& \int_0^{\frac{2\pi b_B^2}{\hbar^2 Q}}{
\delta(D(\varepsilon,K_y))} \nonumber\\
&=& \frac{1}{\pi} \frac{\Theta{\left( |t|^2 - \cos ^2 \phi(\varepsilon) \right)}}{\sqrt{|t|^2 - \cos ^2 \phi(\varepsilon)}}, 
\label{A0}
\end{eqnarray}
%
where $\phi(\varepsilon)=\hbar^2 S(\varepsilon)/(2b_B^2) = \pi \varepsilon^2 /(2 v_F^2 b_B^2) = \pi \varepsilon / (2\hbar \omega_c(\varepsilon))$ and $\Theta(..)$ is the Heaviside theta function. Corrections are neglected in our consideration. Finally, the considered integration  over $K_y$ evaluates to $\int_0^{K_y^m}\delta(D(\varepsilon,K_y)) \approx A_0(\varepsilon)K_y^m$.
Maximal value of conserved momentum $K_y^m$ is determined by the standard condition in the Landau gauge that the electron wave package centred at $x_0$ lies within the sample, i.e. $0 < x_0 < L_x$, yielding $K_y^m=\tfrac{1}{\hbar}eBL_x$.
%
%
Taking it all together, we obtain the temperature-dependent expression for magnetoconductivity
%
\begin{eqnarray}
\sigma_{xx} = -\frac{\zeta}{2\pi^2 \hbar^2 \tau_0 v_F^4 B^2} \int \mathrm{d}\varepsilon \frac{\mathrm{d}f(\varepsilon)}{\mathrm{d}\varepsilon} |\varepsilon|^3 \frac{\Big| \sin \phi(\varepsilon) \Big|}{\sqrt{|t|^2-\cos^2 \phi(\varepsilon)}} \nonumber\\
\times \Theta \left[ |t|^2-\cos^2 \phi(\varepsilon) \right]. \nonumber\\
\label{Sigma_xx_full}
\end{eqnarray}
%

In the absence of magnetic breakdown, i.e. $|t|=1$, the oscillating terms in Eq. (\ref{Sigma_xx_full}) reduce to 1.
The zero-temperature result, with $\mathrm{d}f(\varepsilon)/\mathrm{d}\varepsilon = -\delta(\varepsilon-\varepsilon_F)$ and Sommerfeld correction of the order of $(k_BT/\varepsilon_F)^2$ neglected, reads 
%
\begin{eqnarray}
\sigma_{xx} = \frac{m_F^* \, n_0}{\tau_0 B^2}, 
\label{Sigma_xx_t=1}
\end{eqnarray}
%
where $m_F^* \equiv m^*(\varepsilon_F)$ is an effective cyclotron mass and $n_0=2\zeta S_0(\varepsilon_F)/(2\pi)^2$ is surface concentration of carriers (electrons of both spin projections), both taken at the Fermi energy. Magnetoconductivity has $\sim B^{-2}$ dependence and it does not contain the MB tunneling amplitude $t(B)$. It is equal to the non-oscillating classical result for closed orbits at low temperatures \cite{Abrikozov}.

With finite magnetic breakdown in action, i.e. $|t| < 1$, the oscillating terms in Eq. (\ref{Sigma_xx_full}) remain, causing it to oscillate. By Fourier (re)expansion of the oscillating part under the integral and performing the integration in the complex plane, the characteristic exponential factor appears, i.e. $\exp{(-\pi^2 \varepsilon_F k_B T/2v_F^2 b_B^2)}$. The argument of exponential function is equal to $-(\pi^2/2)(k_BT/\hbar \omega_c)$. It defines the temperature scale $\sim \hbar \omega_c$, up to which the oscillations are visible, in the similar way as Lifshitz-Kosevich formula for Shubnikov - de Haas oscillations \cite{Lifshitz}. For temperatures above this scale, oscillations are exponentially suppressed and what remain is the classical result related with the non-oscillating zeroth coefficient in the above-mentioned expansion. For temperatures significantly lower than the mentioned scale, formally taken in the $T=0$ limit again with $\mathrm{d}f(\varepsilon)/\mathrm{d}\varepsilon \approx -\delta(\varepsilon-\varepsilon_F)$, the Eq. (\ref{Sigma_xx_full}) reduces to
%
\begin{eqnarray}
\sigma_{xx} = \frac{m_F^* \, n_0}{\tau_0 B^2} \frac{\Big| \sin \left( \frac{\pi}{2}\frac{\varepsilon_F}{\hbar\omega_c} \right) \Big|}{\sqrt{|t|^2-\cos^2 \left( \frac{\pi}{2}\frac{\varepsilon_F}{\hbar\omega_c} \right)}} \nonumber\\
\times \Theta \left[ |t|^2-\cos^2 \left( \frac{\pi}{2}\frac{\varepsilon_F}{\hbar\omega_c} \right) \right]. 
\label{Sigma_xx_T=0}
\end{eqnarray}
%

\subsection{Diagonal magnetoconductivity $\sigma_{yy}$}

The diagonal magnetoconductivity along $y$-direction (along the open trajectories in reciprocal space, see Fig. \ref{Fig_CaC6structure}c, $k_x$-direction) is determined by the nonvanishing group velocity $v_y$. Therefore, Eq. (\ref{Sigma_general}) can be expressed in the form
%
\begin{eqnarray}
\hspace{-2mm} \sigma_{yy} = -\frac{2\zeta e^2\tau_0}{L_xL_y} \sum_n \frac{L_y}{2\pi} \int \mathrm{d}K_y v_y^2(K_y,\varepsilon_n(K_y)) \, \frac{\mathrm{d}f(\varepsilon)}{\mathrm{d}\varepsilon}\Big|_{\varepsilon_n(K_y)} \nonumber\\
\label{Sigma_yy1}
\end{eqnarray}
%
in which, using procedure elaborated in Ref. \cite{Kaganov-Slutskin}, we express velocity in terms of the dispersion law $D(\varepsilon,K_y)$, i.e.
%
\begin{eqnarray}
v_y=\frac{1}{\hbar}\frac{\partial \varepsilon_n(K_y)}{\partial K_y} = -\frac{1}{\hbar} \frac{\frac{\partial D(\varepsilon,K_y)}{\partial K_y}}{\frac{\partial D(\varepsilon,K_y)}{\partial \varepsilon}}.
\label{vy_group}
\end{eqnarray}
%
Using the same procedure to change the variable $\varepsilon_n \rightarrow \varepsilon$ as in the previous subsection, as well as Eq. (\ref{identity-delta}), Eq. (\ref{Sigma_yy1}) can be expressed as
%
\begin{eqnarray}
\sigma_{yy} = -\frac{\zeta e^2\tau_0}{\pi \hbar^2 L_x} \int \mathrm{d}K_y \int \mathrm{d}\varepsilon \frac{\mathrm{d}f(\varepsilon)}{\mathrm{d}\varepsilon} \frac{\left( \frac{\partial D}{\partial K_y} \Big|_{\varepsilon} \right)^2}{\Big| \frac{\partial D}{\partial \varepsilon} \Big|_{K_y}  \Big|}\delta(D(\varepsilon,K_y)). \nonumber\\
\label{Sigma_yy2}
\end{eqnarray}
%
Taking the partial derivatives Eq. (\ref{D_partial_derivatives}) in the expression above and using Eqs. (\ref{delta-expansion}) and (\ref{A0}), we obtain
%
\begin{eqnarray}
\sigma_{yy} = -\frac{\zeta e^2\tau_0 Q^2 v_F^2}{\pi^3} \int_{-\infty}^{\infty} \mathrm{d}\varepsilon \frac{\mathrm{d}f(\varepsilon)}{\mathrm{d}\varepsilon} \frac{1}{|\varepsilon|} \frac{\sqrt{|t|^2-\cos^2 \phi (\varepsilon)}}{\Big| \sin \phi (\varepsilon) \Big|} \nonumber\\
\times \Theta \left[ |t|^2-\cos^2 \phi (\varepsilon) \right]. \nonumber\\
\label{Sigma_yy_full}
\end{eqnarray}
%

The zero-temperature result in the absence of the MB-induced over-gap electron tunneling ($|t|=1$) reduces to a constant,
%
\begin{eqnarray}
\sigma_{yy} = \frac{\zeta}{\pi^3} \frac{e^2 \tau_0 Q^2}{m_F^*},
\label{Sigma_yy_t=1}
\end{eqnarray}
%
independent of magnetic field, which coincides with the classical result along the open trajectories proportional to the relaxation time $\tau_0$ \cite{Abrikozov}.

The zero-temperature result with finite magnetic breakdown, obtained from Eq. (\ref{Sigma_yy_full}), attains the form
%
\begin{eqnarray}
\sigma_{yy} = \frac{\zeta}{\pi^3} \frac{e^2 \tau_0 Q^2}{m_F^*} \frac{\sqrt{|t|^2-\cos^2 \left( \frac{\pi}{2}\frac{\varepsilon_F}{\hbar\omega_c} \right)}}{\Big| \sin \left( \frac{\pi}{2}\frac{\varepsilon_F}{\hbar\omega_c} \right) \Big|} \nonumber\\
\times \Theta \left[ |t|^2-\cos^2 \left( \frac{\pi}{2}\frac{\varepsilon_F}{\hbar\omega_c} \right) \right]. 
\label{Sigma_yy_T=0}
\end{eqnarray}
%

\subsection{Hall magnetoconductivity $\sigma_{xy}$}

As in the case of $\sigma_{xx}$, due to vanishing group velocity $v_x$, the corresponding contribution must be accounted through the matrix element $\langle \eta |\hat v_x| \eta' \rangle$ which is expressed in terms of the Lorentz force operator proportional to $\dot{\hat{k}}_y$. Expression (\ref{Sigma_general}) in the case of Hall conductivity reduces to
%
\begin{eqnarray}
\sigma_{yx} = &-& \frac{2\zeta e}{L_x L_y B} \sum_{\eta} \left( \hbar \langle \eta \vert \hat{k}_y \hat{v}_y \vert \eta \rangle  \right. \nonumber\\  
&-& \left. \langle \eta \vert \hat{k}_y \vert \eta \rangle \frac{\partial \varepsilon_n(K_y)}{\partial K_y} \right) \, \frac{\mathrm{d}f(\varepsilon)}{\mathrm{d}\varepsilon}\Big|_{\varepsilon_\eta},
\label{Sigma_xy1}
\end{eqnarray}
%
where $|\eta\rangle = |n,K_y\rangle$. Two matrix elements, $\langle \eta \vert \hat{k}_y \hat{v}_y \vert \eta \rangle$ and $\langle \eta \vert \hat{k}_y \vert \eta \rangle$ are evaluated as follows.
The first one reads
\begin{eqnarray}
\langle \eta \vert \hat{k}_y \hat{v}_y \vert \eta \rangle =\frac{L_x}{2\pi}\int_0^Q  \mathrm{d}k_x\left( 
\frac{\vert C_+ \vert ^2 k_y^+ v_y^+}{\vert v_y^+ \vert}+
\frac{\vert C_- \vert ^2 k_y^-v_y^-}{\vert v_y^- \vert}\right). \nonumber\\
\label{Hall_1st_contrib1}
\end{eqnarray}
Since $k_y^-=-k_y^+$ and $\vert v_y^+\vert = \vert v_y^- \vert$, it reduces to
\begin{eqnarray}
\langle \eta \vert \hat{k}_y \hat{v}_y \vert \eta \rangle &=& \frac{L_x}{2\pi} \int_0^Q \mathrm{d}k_x k_y^+(k_x,\varepsilon_n(K_y)) \left( \vert C_+ \vert ^2 + \vert C_- \vert ^2 \right) \nonumber \\
&=& \frac{L_x}{2\pi} \frac{S_0(\varepsilon_n(K_y))}{2} \left( \vert C_+(K_y) \vert ^2 + \vert C_-(K_y) \vert ^2 \right).\nonumber \\
\label{Hall_1st_contrib2}
\end{eqnarray}
The second term reads
\begin{eqnarray}
\langle \eta \vert \hat{k}_y \vert \eta \rangle &=& \frac{L_x}{2\pi}\int_0^Q  \mathrm{d}k_x \left( \frac{\vert C_+ \vert ^2 k_y^+}{\vert v_y^+ \vert}+ \frac{\vert C_- \vert ^2 k_y^-}{\vert v_y^- \vert}\right) \nonumber \\
&=&  \frac{L_x}{2\pi} \int_0^Q \mathrm{d}k_x \frac{k_y^+}{v_y^+} \left(\vert C_+ \vert ^2 - \vert C_- \vert ^2\right),
\label{Hall_2nd_contrib1}
\end{eqnarray}
where the negative sign in front of $\vert C_- \vert ^2$ appears due to $k_y^-=-k_y^+$.
Using $v_y = \hbar\tfrac{\partial \varepsilon}{\partial k_y}=\tfrac{v_F^2}{\varepsilon}\hbar k_y$, and $\tfrac{k_y^+}{v_y^+}=\tfrac{\varepsilon}{\hbar v_F^2}$, we obtain
\begin{eqnarray}
\langle \eta \vert \hat{k}_y \vert \eta \rangle &=& \frac{L_x}{2\pi}\int_0^Q  \mathrm{d}k_x \frac{\varepsilon_n}{\hbar v_F^2} \left(\vert C_+\vert ^2 - \vert C_-\vert ^2\right) \nonumber\\
&=& \frac{L_x Q}{2\pi} \frac{\varepsilon_n}{\hbar v_F^2} \left(\vert C_+ (K_y) \vert ^2 - \vert C_-(K_y) \vert ^2\right).
\label{Hall_2nd_contrib2}
\end{eqnarray}
Similarly to expression (\ref{NormalizationSum_II}), for $\vert C_+ \vert ^2 + \vert C_-\vert^2$, using Eqs. (\ref{CoefficientsRelation}) and (\ref{NormalizationSum_II}), one can obtain an expression for $\vert C_+ \vert ^2 - \vert C_-\vert^2$, i.e.  
\begin{eqnarray}
\vert C_+ \vert ^2  -   \vert C_- \vert ^2 = \frac{2\pi\hbar}{eBL_x} \frac{2}{T(\varepsilon)} |t| \frac{|t|-\cos(\phi_1 + \phi_2)}{1-|t|\cos(\phi_1 + \phi_2)},\nonumber\\
\label{integral_razlike_C-ova}
\end{eqnarray}
where $\phi_1 \equiv \frac{\pi \varepsilon^2}{2v_F^2 b_B^2}$ and $\phi_2 \equiv \frac{\hbar^2 QK_y}{b_B^2}$.

Now we can evaluate Hall conductivity consisting of two contributions, i.e. $\sigma_{yx}=\sigma^{\mathrm{I}}_{yx}+\sigma^{\mathrm{II}}_{yx}$. The first contribution $\sigma^{\mathrm{I}}_{yx}$, corresponding to Eq. (\ref{Hall_1st_contrib2}), is evaluated in the analogous way as $\sigma_{xx}$ by changing the variable $\varepsilon_n \rightarrow \varepsilon$ and using Eqs. (\ref{D_partial_derivatives}), (\ref{identity-delta}), (\ref{delta-expansion}) and (\ref{A0}), yielding
%
\begin{eqnarray}
\sigma^{\mathrm{I}}_{yx} = -\frac{\zeta e}{2\pi \hbar^2 v_F^2 B} \int \mathrm{d}\varepsilon \frac{\mathrm{d}f(\varepsilon)}{\mathrm{d}\varepsilon} \varepsilon^2 \frac{\Big| \sin \phi(\varepsilon) \Big|}{\sqrt{|t|^2-\cos^2 \phi(\varepsilon)}} \nonumber\\
\times \Theta \left[ |t|^2-\cos^2 \phi(\varepsilon) \right]. \nonumber\\
\label{Sigma_yx_I_full}
\end{eqnarray}
%

The second contribution $\sigma^{\mathrm{II}}_{yx}$, corresponding to Eq. (\ref{Hall_2nd_contrib2}), contains the expression
\begin{eqnarray}
-\frac{\partial \varepsilon_n}{\partial K_y} \langle \eta \vert k_y \vert \eta \rangle = \frac{\varepsilon_n Q}{\hbar v_F^2}\frac{\partial D}{\partial K_y}\left(\vert C_+ \vert ^2 - \vert C_- \vert ^2 \right)\delta(D), \,\,\,\,\,\,\,\,
\label{intermediate}
\end{eqnarray}
where $D$ is defined in expression (\ref{D}). Eq. (\ref{intermediate}) is a fast oscillating function containing the expression
\begin{eqnarray}
F(\phi_1, \phi_2) = \frac{|t|-\cos \left(\phi_1+ \phi_2 \right)}{1-|t|\cos\left(\phi_1+ \phi_2 \right)}\sin\phi_2,
\label{eq.f}
\end{eqnarray}
which is periodic with $2\pi$ in $\phi_1$ and $\phi_2$.
We expand expression (\ref{eq.f}) in the Fourier series, i.e.
\begin{eqnarray}
F(\phi_1, \phi_2) = \sum_{s,l} F_{s,l} \exp{\left[ i \left( s\phi_1 + l\phi_2\right) \right]},
\label{FourierF}
\end{eqnarray}
where $F_{s,l}$ are expansion coefficients for all integer $s$ and $l$. Again, we keep just the main contribution $F_{0,0} \equiv \bar{F}$ (with $s=l=0$), i.e. $F(\phi_1, \phi_2) \approx \bar{F}$, where
\begin{eqnarray}
\bar{F} &=& \int_{-\pi}^{\pi}\frac{\mathrm{d}\phi_1}{2\pi}\int_{-\pi}^{\pi}\frac{\mathrm{d}\phi_2}{2\pi}
\frac{|t|-\cos \phi_1 \cos \phi_2  + \sin \phi_1 \sin\phi_2}{1-|t|(\cos \phi_1 \cos \phi_2- \sin \phi_1 \sin\phi_2 )} \nonumber\\ &\times & \frac{\sin\phi_2}{\sqrt{1-|t|^2\cos^2 \phi_2}} \, \delta \left( \cos\phi_1 - |t|\cos \phi_2\right), 
\label{fsrednje}
\end{eqnarray}
neglecting the fast-oscillating corrections with $s\ne 0$, $l \ne 0$.
The $\delta$-function inside the integral in Eq. (\ref{fsrednje}) is evaluated as a sum of $\delta$-functions over all zeroes in the domain of integration, finally yielding $\bar{F}=0$.
Therefore, the second contribution to the main part of Hall conductivity evaluates to $\sigma^{\mathrm{II}}_{yx}\approx 0$ leaving the term $\sigma^{\mathrm{I}}_{yx}$ as the contributing one.

The zero-temperature limit in the absence of magnetic breakdown ($|t|=1$) yields the result
\begin{eqnarray}
\sigma_{xy}=-\sigma_{yx}=-\frac{en_0}{B}
\label{Sigma_yx_t=1}
\end{eqnarray}
written in terms of the total 2D electron concentration $n_0=2\zeta S_0(\varepsilon_F)/(2\pi)^2$.  This result corresponds to the classical one in a strong field \cite{Abrikozov}.

The zero-temperature result with finite magnetic breakdown, obtained from Eq. (\ref{Sigma_yy_full}), attains the form
%
\begin{eqnarray}
\sigma_{xy}=-\sigma_{yx}=-\frac{en_0}{B} \frac{\Big| \sin \left( \frac{\pi}{2}\frac{\varepsilon_F}{\hbar\omega_c} \right) \Big|}{\sqrt{|t|^2-\cos^2 \left( \frac{\pi}{2}\frac{\varepsilon_F}{\hbar\omega_c} \right)}} \nonumber\\
\times \Theta \left[ |t|^2-\cos^2 \left( \frac{\pi}{2}\frac{\varepsilon_F}{\hbar\omega_c} \right) \right].
\label{Sigma_yx_T=0}
\end{eqnarray}
%

\bigskip

\begin{figure*}
\centerline{\includegraphics[width=1.75\columnwidth]{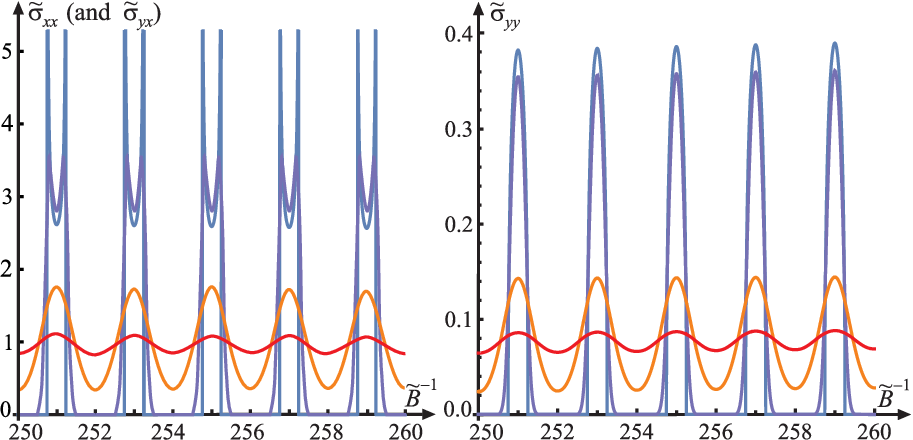}}
\caption{Quantum oscillations of magnetoconductivity components $\sigma_{ij}$ in the presence of finite magnetic breakdown, depending on magnetic field and temperature. Components $\widetilde{\sigma}_{xx}$, $\widetilde{\sigma}_{yy}$ and $\widetilde{\sigma}_{yx}$ are plotted according to expressions (\ref{Sigma_xx_full}), (\ref{Sigma_yy_full}) and (\ref{Sigma_yx_I_full}) scaled to $n_0 m_F^* / (\tau_0 B^2)$, $\zeta e^2 \tau_0 Q^2 / (\pi^3 m_F^*)$ and $e n_0 / B$, respectively in each expression. Graphs represent (scaled) energy integrals of oscillations-generating functions, embedded with the derivative of Fermi function, vs. inverse magnetic field $B^{-1}$ (scaled to magnetic energy i.e. $\widetilde{B} \equiv \hbar\omega_c/\varepsilon_F$). Graphs for $\widetilde{\sigma}_{xx}$ and $\widetilde{\sigma}_{yx}$ are on this scale indistinguishable to the eye, thus are presented on the same figure, while both figures present a characteristic pattern of oscillations with respect to the characteristic scale (period determined by the area of electron trajectory $S_0(\varepsilon_F)$ vs. $B^{-1}$). The temperatures (scaled to $k_B/\varepsilon_F$) in figures are: 0 (blue), 0.0001 (purple), 0.0005 (orange), 0.001 (red). The low-temperature "gaps" of zero conductivity are closed more and more at higher temperatures until the oscillations get suppressed at temperatures higher that $\hbar \omega_c$ scale. The energy gap parameter in spectrum is everywhere $\Delta/\varepsilon_F = 0.01$ i.e. of the characteristic order of magnitude of $10^2$ K.} 
\label{Fig_Sigma_osc}
\end{figure*}

\begin{figure}
\centerline{\includegraphics[width=.8\columnwidth]{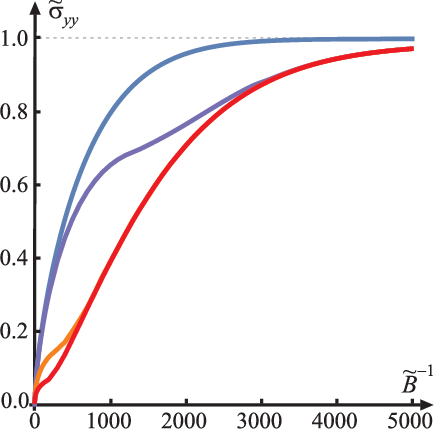}}
\caption{The upper envelope of fast-oscillating $\sigma_{yy}$ vs. $B^{-1}$ at different temperatures. $\widetilde{\sigma}_{yy}$ is scaled to $\zeta e^2 \tau_0 Q^2 / (\pi^3 m_F^*)$, while $B$ is scaled as $\widetilde{B} \equiv \hbar\omega_c/\varepsilon_F$. The temperatures (scaled to $k_B/\varepsilon_F$) in figure are: 0 (blue), 0.0001 (purple), 0.0005 (orange), 0.001 (red). The energy gap parameter in spectrum is $\Delta/\varepsilon_F = 0.01$.} 
\label{Fig_Sigma_yy}
\end{figure}

Both in expressions for magnetoconductivity components in the presence of finite magnetic breakdown at finite temperature, Eqs. (\ref{Sigma_xx_full}), (\ref{Sigma_yy_full}), (\ref{Sigma_yx_I_full}), and those at $T=0$, Eqs. (\ref{Sigma_xx_T=0}), (\ref{Sigma_yy_T=0}), (\ref{Sigma_yx_T=0}), the Heaviside theta function insures that argument of oscillating function inside expressions is within the (magnetic) band. Simple analysis of the zero-temperature results shows zero values of magnetoconductivity at fields for which $\varepsilon_F/\hbar \omega_c$ is equal to an even integer. Around these values there are finite "gaps" of zero-conductivity, determined by the afore-mentioned Heaviside function, which correspond to gaps in the energy spectrum between magnetic bands (see Fig. \ref{Fig_Spectrum}). The widths of these gaps depend on magnetic field also through the dependence of $t(B)$ (see Fig. \ref{Fig_t2}). Finite values of magnetoconductivity outside of these gaps, appearing periodically with $B^{-1}$ with period proportional to $S_0(\varepsilon_F)$, essentially represent the onset of quantum oscillations, the sharper, the temperature is lower (see Fig. \ref{Fig_Sigma_osc}). 
Temperature is one of usual channels to "smooth" the oscillations, appearing with oscillating integrand under the integral in the temperature-dependent expressions. This particular channel is present in our model while the other one, the relaxation time (or analogous self-energy), is ruled out by our starting condition $\omega_c \gg \tau_0^{-1}$ required to achieve the coherent MB. Therefore, in order to observe these oscillations, one needs to provide clean samples and low temperatures.\
%
%
It is evident that in both, temperature-dependent and zero-temperature expressions for magnetoconductivity components $\sigma_{ij}$, the finite MB effect modifies otherwise non-oscillating main ("classical") contribution in terms of emerging quantum oscillations. Those oscillations do not appear as the additive small correction to the classical part, but rather as inherent to it, forcing it to oscillate from zero to its maximal value (at low enough temperatures).
As mentioned earlier, apart from the onset of oscillations, the MB affects the magnetoconductivity through the field-dependent MB transmission probability amplitude $t(B)$, which modifies the magnetic band width and affects the amplitude of quantum oscillations at different fields. We illustrate it in Fig. \ref{Fig_Sigma_yy} on example of $\sigma_{yy}$. Its "classical" low field value, when $|t| \rightarrow 1$ and electron trajectories are open, is otherwise constant. With increasing field the over-gap MB tunneling increases ($|t|$ gets smaller than 1), reconstructing the closed electron trajectories and effectively increasing electron localization. Emerging oscillations drop in amplitude with increasing field as $|t|$ decreases. We have to mention that, using presented results, one cannot analytically perform a crossover to $|t|=0$ limit (closed orbits) due to entirely different structure of starting velocity operators to be used in magnetoconductivity calculation in that case. Besides that, we repeat that validity of our description holds until magnetic band width is larger than electron level broadening due to scattering on impurities, which is not the case in the limit of very narrow magnetic bands tending towards Landau levels ($|t|=0$).
%
%

\section{Conclusions}

What makes the CaC$_6$ system in the CDW ground state gainful to study effects of magnetic breakdown is an exact match of  natural scales in the problem required for them to be pronounced. In particular, the observed uniaxial CDW \cite{Rahnejat,Shimizu} reconstructs the closed Fermi pockets into open sheets with characteristic spacings between them at the Brillouin zone edges approaching values determined by the gap parameter of the order of $10^2$K \cite{Petra}  (experimental values: critical CDW temperature is 250K, pseudogap width from differential conductivity is 475 mV \cite{Rahnejat}). Those exactly match the MB requirement for magnetic fields of the order of $B \sim 10\,$T \cite{Kaganov-Slutskin}. In the samples clean enough to provide mean free path significantly longer comparing to the Larmor radius in those fields, with absence of dislocation fields, and temperature low enough to neglect relaxation channels other than scattering of carriers on impurities, we predict properties of the magnetoconductivity tensor, with the influence of magnetic breakdown to the main, so-called its "classical part" in focus, and neglecting additive corrections to it.\

Magnetoconductivity is calculated within quantum density matrix approach and semiclassical approximation based on the Lifshitz-Onsager Hamiltonian, using specific technique developed for magnetic breakdown \cite{Kaganov-Slutskin}. Under the circumstances, electron spectrum consists of so-called magnetic bands, wider than average level broadening due to impurity scattering. 
It means that the limit of closed electron trajectories, to which the system would tend in the limit of extreme magnetic breakdown (which would restore closed trajectories from the open ones and reduce the picture to the Landau quantization physics), is not covered within our description of magnetoconductivity. In the absence of magnetic breakdown, i.e. when the transmission probability between neighboring cells $|t|^2 \rightarrow 1$ at low fields, preserving the picture of open trajectories, the components of magnetoconductivity tensor attain their classical values \cite{Abrikozov}. These zero-temperature values are: $\sigma_{xx}\sim B^{-2}$ for component along the CDW crests which is perpendicular to the open semiclassical electron trajectories in reciprocal space, $\sigma_{yy}\sim \mathrm{const}.$ for component along the open trajectories, and $\sigma_{xy}\sim B^{-1}$ for Hall conductivity.
On the other hand, in the regime of finite magnetic breakdown $|t|^2 < 1$, all three magnetoconductivity components start to manifest strong quantum oscillations vs. inverse magnetic field at low temperatures. They are most pronounced at zero-temperature and get gradually "smoothed out" as temperature increases, finally being exponentially suppressed at temperatures higher than magnetic energy scale $\hbar \omega_c$. 
The onset of oscillations is a feature of coherent magnetic breakdown, in which large number of electrons takes part. It is due to the interference of huge semiclassical phases (comparing to the characteristic magnetic scale $\sim b_B^2$) characterizing the semiclassical wave functions.
In our approach we kept the first order term in that interference, neglecting terms of higher orders.
These oscillations are not just a mere additive correction appearing on top of the classical part, which start to oscillate from zero to maximal value, they are an {\it inherent} part of it thus turning it to essentially non-classical.
These are not standard Shubnikov - de Haas (SdH) oscillations, appearing in systems with closed Fermi surface due to Landau quantization where the onset of oscillations lies in modifications of DOS that pass through the Fermi level as the field is changed. It generates an additive oscillating correction to the principal classical result (at low enough fields before the so-called ultra-quantum limit takes place). In CaC$_6$ system under CDW, with open sheets of the Fermi surface, standard SdH type of oscillations does not exist, and the only oscillatory behavior may appear due to magnetic breakdown.\

What is usually measured in experiments is the magnetoresistivity related to magnetoconductivity by inversion of its tensor, i.e. {\boldmath $\rho$}={\boldmath $\sigma$}$^{-1}$. Although the low-temperature oscillatory behavior is of the same manner in the sense of frequency vs. $B^{-1}$, the classical field-dependent "envelope" of each magnetoresistivity component depends on all magnetoconductivity components, i.e. $\rho_{xx/yy}=\sigma_{yy/xx}/(\sigma_{xx}\sigma_{yy}+\sigma_{xy}^2)$. Taking the low-temperature values of magnetoconductivity components from expressions (\ref{Sigma_xx_t=1}), (\ref{Sigma_yy_t=1}) and (\ref{Sigma_yx_t=1}) we obtain expressions for magnetoresistivity envelopes $\rho_{xx}=\zeta \tau_0 Q^2 B^2 / (\pi^3 n_0^2 m_F^* \Upsilon) \sim B^2$, $\rho_{yy}= m_F^* / (e^2 \tau_0 n_0 \Upsilon) \sim \mathrm{const}$, where $\Upsilon \equiv 1+\zeta Q^2/(\pi^3 n_0)$. In contrast to systems with closed Fermi surface, in which there appears characteristic crossover from low-field $\sim B^2$ behavior into high-field saturation to constant magnetoresistivity, here the $\sim B^2$ behavior of $\rho_{xx}$ persists (within validity of the semiclassical approximation).\

To our knowledge, experimental values for magnetoconductivity of CaC$_6$ in the CDW groundstate (open Fermi surface) are not available so far. There are, however, measurements of magnetoresistance in CaC$_6$ in normal groundstate (absence of the CDW) with closed Fermi surface \cite{Mu}. The magnetoresistance shows quadratic in field dependence as expected in low, which crosses over to linear in strong fields instead of saturation which would be expected metallic behavior. There is a rather old paper by A. A. Abrikosov \cite{Abrikozov2} which relates linear in field magnetoresistance with linearity of electron spectrum, but in the strong field vs. low doped system  with closed Fermi surface and Dirac point in spectrum, in which only one Landau level (the first one) one is engaged. In that respect, CaC$_6$ with chemical doping of 0.2 electrons per carbon atom and consequently the Fermi energy of the order of electronvolt is quite highly doped comparing to magnetic fields of several Tesla where linear in field behavior is observed. Although we took the linearity of electron spectrum into account and our result for $\sigma_{xx}$ should correspond to one for the system with closed Fermi surface, it needs to be stressed that the semiclassical picture is valid in the limit of large values of (occupied) Landau band indices. It is exactly opposite to Abrikosov's limit and corresponds more to the limit in which the observed magnetoresistivity is quadratic in field. In that sense explanation of observed linear in field magnetoresistance remains an open question.

\section*{Acknowledgements}

This work was supported by the QuantiXLie Centre of Excellence, a project co-financed by the Croatian
Government and European Union through the European Regional Development Fund - the Competitiveness and Cohesion Operational Programme (Grant PK.1.1.02). The authors are grateful to dr. I. Smoli\'{c} for constructive discussions.

\end{document}